\documentstyle[11pt]{article}
\input amssymb.sty

\title{{\it Probabilistic Knowledge} as {\it Objective Knowledge}\\ in Quantum Mechanics:\\ {\it Potential Powers} Instead of {\it Actual Properties}.}

\author{{\sc Christian de Ronde}\thanks{Fellow Researcher of the Consejo
Nacional de Investigaciones Cient\'{\i}ficas y T\'ecnicas.}}
\date{}

\begin{document}

\bibliographystyle{plain}
\maketitle

\begin{center}
\begin{small}
Philosophy Institute ``Dr. A. Korn" \\ 
Buenos Aires University, CONICET - Argentina \\
Center Leo Apostel and Foundations of  the Exact Sciences\\
Brussels Free University - Belgium \\
\end{small}
\end{center}

\begin{abstract}
\noindent In classical physics, probabilistic or statistical knowledge has been always related to ignorance or inaccurate subjective knowledge about an actual state of affairs. This idea has been extended to quantum mechanics through a completely incoherent interpretation of the Fermi-Dirac and Bose-Einstein statistics in terms of ``strange'' quantum particles. This interpretation, naturalized through a widespread ``way of speaking'' in the physics community, contradicts Born's physical account of $\Psi$ as a ``probability wave'' which provides statistical information about outcomes that, in fact, cannot be interpreted in terms of `ignorance about an actual state of affairs'. In the present paper we discuss how the metaphysics of actuality has played an essential role in limiting the possibilities of understating things differently. We propose instead a metaphysical scheme in terms of {\it powers} with definite {\it potentia} which allows us to consider quantum probability in a new light, namely, as providing objective knowledge about a potential state of affairs.
\end{abstract}
\begin{small}


\end{small}

\newtheorem{theo}{Theorem}[section]

\newtheorem{definition}[theo]{Definition}

\newtheorem{lem}[theo]{Lemma}

\newtheorem{met}[theo]{Method}

\newtheorem{prop}[theo]{Proposition}

\newtheorem{coro}[theo]{Corollary}

\newtheorem{exam}[theo]{Example}

\newtheorem{rema}[theo]{Remark}{\hspace*{4mm}}

\newtheorem{example}[theo]{Example}

\newcommand{\proof}{\noindent {\em Proof:\/}{\hspace*{4mm}}}

\newcommand{\qed}{\hfill$\Box$}

\newcommand{\ninv}{\mathord{\sim}} %involutive negation

\newtheorem{postulate}[theo]{Postulate}

\vspace{0.5cm}

\section{Reality, Metaphysics and Knowledge}

The term ``metaphysics'' comprises a series of many different definitions. The most known is the one proposed by Aristotle who defined it as a theory of ``being {\it qua} being'' [Met. 1003a20], a theory about what it means or implies to ``be'' in its different senses. Ever since, metaphysics and its problems have remained at the center of debates in western philosophical thought. But while for some, it is considered as a supreme form of knowledge, for others, it remains an occupation constituted by unfruitful discussions and pseudoproblems.  One of the very first metaphysical problems is the one known by the name of ``the problem of motion'', a problem which goes back to pre-socratic philosophy and was brought to us by Plato and Aristotle. Plato and Aristotle created an opposition between Heraclitus, who embraced the doctrine of permanent motion and becoming in the world; and Parmenides, who taught  the non-existence of motion and change in reality, reality being absolutely one and determined.

\begin{quotation}
\noindent {\small ``The contradicting conclusions deriving from pre-Socratic philosophy were of a major concern to Plato and Aristotle, because they challenged the existence of truth and certainty about the world and therefore about the actions of human beings in it. This uncertainty had given rise to a philosophical discipline, Sophism, that simply denied any relation between reality and what we say about it ([45], {\it Theaetetus}, 42, 152(d,e)). Its subjectivism stems from a radical empiricism, which holds that things are for me as I perceive them. But since reality as we perceive it is always in a process of permanent change this implies, as Plato points out in the {\it Theaetetus}, also the non-existence of stable, individual things in the world.'' \cite[p. 164]{VerelstCoecke}}
\end{quotation} 

Metaphysics introduced the fundamental idea of acquiring knowledge of reality and existence through a set of specific principles. One of the first such schemes, which still today plays a major role in our understanding of the world around us is that proposed by Aristotle through his logical and ontological principles: the Principle of Existence (PE), the Principle of Non-Contradiction (PNC) and the of Principle Identity (PI). 

\begin{quotation}
\noindent {\small ``The three fundamental principles of classical (Aristotelian) logic: the existence of objects of knowledge, the principle of contradiction and the principle of identity, all correspond to a fundamental aspect of his ontology. This is exemplified in the three possible usages of the verb Òto beÓ: existential, predicative, and identical. The Aristotelian syllogism always starts with the affirmation of existence: something is. The principle of contradiction then concerns the way one can speak (predicate) validly about this existing object, i.e. about the true and falsehood of its having properties, not about its being in existence. The principle of identity states that the entity is identical to itself at any moment (a=a), thus granting the stability necessary to name (identify) it.'' \cite[p. 169]{VerelstCoecke}}
\end{quotation}

\noindent Aristotle had developed a metaphysical scheme in which, through the notions of {\it actuality} and {\it potentiality}, he was able to articulate both the Heraclitean and the Eleatic schools. On the one hand, potentiality contained the undetermined, contradictory and non-individual realm of existence, on the other, the mode of being of actuality was determined through the PE, PNC and PI. Through these principles the notion of entity was capable of unifying, of totalizing in terms of a ``sameness'', creating certain stability for knowledge to be possible. Although Aristotle claimed that Being is said in many ways presenting at first both actual and potential realms as ontologically equivalent, from chapter 6 of book $\Theta$ of {\it Metaphysics}, he seems to place actuality in the central axis of his architectonic, relegating potentiality to a mere supplementary role.\footnote{Aristotle argues: ``We have distinguished the various senses of `prior', and it is clear that actuality is prior to potentiality. [...] For the action is the end, and the actuality is the action. Therefore even the word `actuality' is derived from `action', and points to the fulfillment.'' [1050a17-1050a23] Aristotle then continues to provide arguments in this line which show ``[t]hat the good actuality is better and more valuable than the good potentiality.'' [1051a4-1051a17]} 

Both actuality and potentiality were part of a metaphysical representation and understood as characterizing modes of existence independent of observation. This is the way through which metaphysical thought was able to go beyond the  {\it hic et nunc}, creating a world beyond the world, a world of concepts and representations. Such representation or transcendent description of the world is considered by many as the origin of metaphysical thought itself.\footnote{The need of metaphysical principles in order to account for physical experience has been beautifully exposed by Borges in a story called {\it Funes the Memorious}.} And this is the reason why, as noticed by Edwin Burtt \cite[p. 224]{Burtt03}: ``[...] there is no escape from metaphysics, that is, from the final implications of any proposition or set of propositions. The only way to avoid becoming a metaphysician is to say nothing.''

\section{Classical Physics: Actual Properties and States of Affairs}

The importance of potentiality, which was first placed by Aristotle in equal footing to actuality as a mode of existence, was soon diminished in the history of western thought. As we have seen above, it could be argued that the seed of this move was already present in the Aristotelian architectonic, whose focus was clearly placed in the actual realm. The realm of potentiality, as a different (ontological) mode of the being was neglected becoming not more than mere (logical) {\it possibility}, a teleological process of fulfillment. In relation to the development of physics, the focus and preeminence was also given to actuality. The XVII century division between {\it res cogitans} and {\it res extensa} played in this respect an important role separating very clearly the realms of actuality and potentiality. The philosophy which was developed after Descartes kept `res cogitans' (thought) and `res extensa' (entities as acquired by the senses) as separated realms.\footnote{While `res cogitans', the soul, was related to the {\it indefinite} realm of potentiality, `res extensa', i.e. the entities as characterized by the principles of logic, related to the actual.} As remarked by Heisenberg \cite[p. 73]{Heis58}: ``Descartes knew the undisputable necessity of the connection, but philosophy and natural science in the following period developed on the basis of the polarity between the `res cogitans' and the `res extensa', and natural science concentrated its interest on the `res extensa'." This materialistic conception of science based itself on the main idea that extended things exist as being definite, that is, in the actual realm of existence. With modern science the actualist Megarian path was recovered and potentiality dismissed as a problematic and unwanted guest. The transformation from medieval to modern science coincides with the abolition of Aristotelian hilemorphic metaphysical scheme ---in terms of potentiality and actuality--- as the foundation of knowledge. However, the basic structure of his metaphysical scheme and his logic still remained the basis for correct reasoning. As Verelst and Coecke remark:

\begin{quotation}
\noindent {\small ``Dropping Aristotelian metaphysics, while at the same time
continuing to use Aristotelian logic as an empty `reasoning
apparatus' implies therefore loosing the possibility to account for
change and motion in whatever description of the world that is based
on it. The fact that Aristotelian logic transformed during the
twentieth century into different formal, axiomatic logical systems
used in today's philosophy and science doesn't really matter,
because the fundamental principle, and therefore the fundamental
ontology, remained the same ([40], p. xix). This `emptied' logic
actually contains an Eleatic ontology, that allows only for static
descriptions of the world."
\cite[p. 169]{VerelstCoecke}}
\end{quotation}

It was Isaac Newton who was able to translate into a closed
mathematical formalism both, the ontological presuppositions present
in Aristotelian (Eleatic) logic and the materialistic ideal of `res
extensa' ---with actuality as its mode of existence. 
In classical mechanics the representation of the state of the
physical system is given by a point in phase space $\Gamma$ and the
physical magnitudes are represented by real functions over $\Gamma$.
These functions commute between each other and can be
interpreted as possessing definite values independently of
measurement, i.e. each function can be interpreted as being actual.
The term `actual' refers here to {\it preexistence} (within the
transcendent representation) and not to {\it hic et
nunc} observation. Every physical system may be described
exclusively by means of its actual properties. The change of the system may be
described by the change of its actual properties. Potential or
possible properties are considered as the points to which the system
might arrive in a future instant of time. As also noted by Dieks:

\begin{quotation}
\noindent {\small ``In classical physics the most fundamental description of a physical system (a point in phase space) reflects only the actual, and nothing that is merely
possible. It is true that sometimes states involving probabilities
occur in classical physics: think of the probability distributions
$\rho$ in statistical mechanics. But the occurrence of possibilities
in such cases merely reflects our ignorance about what is actual.
The statistical states do not correspond to features of the actual
system (unlike the case of the quantum mechanical superpositions),
but quantify our lack of knowledge of those actual features.''
\cite[p. 124]{Dieks10}}
\end{quotation}

\noindent Classical mechanics tells us via the equation of motion how the state of the system moves along the curve determined by the initial conditions in $\Gamma$ and thus, as any mechanical property may be expressed in terms of $\Gamma$'s variables, how all of them evolve. Moreover, the structure in which actual properties may be organized is the (Boolean) algebra of classical logic.

\section{Physical Probability and Subjective Ignorance}

We believe that a realist coherent interpretation of Quantum Mechanics (QM) should be capable of providing an internal understanding of its physical concepts. Since Born's 1926 interpretation of the quantum wave function $\Psi$, probability has become one of the key notions in the description of quantum phenomena. But the difficulties to interpret quantum probability were already explicit in Born's original paper. 

\begin{quotation}
\noindent {\small ``Schr\"{o}dinger's quantum mechanics [therefore] gives quite a definite answer to the question of the effect of the collision; but there is no question of any causal description. One gets no answer to the question, `what is the state after the collision' but only to the question, `how probable is a specified outcome of the collision'.

Here the whole problem of determinism comes up. From the standpoint of our quantum mechanics there is no quantity which in any individual case causally fixes the consequence of the collision; but also experimentally we have so far no reason to believe that there are some inner properties of the atom which condition a definite outcome for the collision. [...] I myself am inclined to give up determinism in the world of the atoms. But that is a philosophical question for which physical arguments alone are not decisive.'' \cite[p. 57]{WZ}}
\end{quotation}

\noindent In his paper Born formulated the now-standard interpretation of $\psi (x)$ as encoding a probability density function for a certain particle to be found at a given region. The wave function is a complex-valued function of a continuous variable. For a state $\psi$, the associated probability function is $\psi^{*} \psi$, which is equal to $|\psi (x)|^2$. If $|\psi (x)|^2$ has a finite integral over the whole of three-dimensional space, then it is possible to choose a normalizing constant. The probability that a particle is within a particular region V is the integral over V of $|\psi (x)|^2$. However, even though this interpretation worked fairly well, it soon became evident that the concept of probability in the new theory departed from the physical notion considered in classical statistical mechanics as {\it lack of knowledge} about a preexistent (actual) state of affairs described in terms of definite valued properties.

In the history of physics the development of probability took place through a concrete physical problem and has a long history which goes back to the 18th century. The physical problem with which probability dealt was the problem of characterizing a state of affairs of which there is an incomplete knowledge. Or in other words, ``gambling". This physical problem was connected later on to a mathematical theory developed by Laplace and others. But it was only after Kolmogorov that this mathematical theory found a closed set of axioms \cite{Kolmogorov}. Although there are still today many interpretational problems regarding the physical understanding of classical probability, when a realist physicist talks about probability in statistical mechanics he is discussing about the (average values of) properties of an uncertain ---but existent---  state of affairs.\footnote{In this respect it is important to remark that the orthodox interpretation of probability in terms of relative frequencies, although provides a conceptual framework to relate to measurement outcomes, refers to `events' and to `properties of a system'; in this sense it is not necessarily linked to a realistic physical representation but rather supports an empiricist account of the observed measurement results.} This is why the problem to determine a definite state of affairs in QM ---the sets of definite valued properties which characterize the quantum system--- poses also problems to the interpretation of probability and possibility within the theory itself. As noticed by Schr\"odinger in a letter to Einstein:

\begin{quotation}
\noindent{\small ``It seems to me that the concept of probability is terribly
mishandled these days. Probability surely has as its substance a
statement as to whether something {\small {\it is}} or {\small {\it
is not}} the case ---an uncertain statement, to be sure. But
nevertheless it has meaning only if one is indeed convinced that the
something in question quite definitely {\small {\it is}} or {\small
{\it is not}} the case. A probabilistic assertion presupposes the
full reality of its subject.'' \cite[p. 115]{Bub97}}
\end{quotation}

We understand mathematics, contrary to physics, as a non-representational discipline which respects no metaphysical nor empirical limits whatsoever. The mathematician does not need to constrain himself to any sort of metaphysical principles but only to the internal structure of the mathematical theory itself. `Probability' is regarded by the mathematician as a `theory of mathematics' and in this sense departs from any conceptual physical understanding which relates the formal structure to the world around us. A mathematician thinks of a probability model as the set of axioms which fit a mathematical structure and wonders about the internal consistency rather than about how this structure relates and can be interpreted in relation to experience and physical reality. As noticed by Hans Primas:

\begin{quotation}
\noindent {\small ``Mathematical probability theory is just a branch of pure
mathematics, based on some axioms devoid of any interpretation. In
this framework, the concepts `probability', `independence', etc. are
conceptually unexplained notions, they have a purely mathematical
meaning. While there is a widespread agreement concerning the
essential features of the calculus of probability, there are widely
diverging opinions what the referent of mathematical probability
theory is.''  \cite[p. 582]{Primas99}}\end{quotation}

\noindent The important point is that when a mathematician and a physicist talk about `probability' they need not refer to the {\it same} concept. While for the mathematician the question of the relation between the mathematical structure of probability and experience plays no significant role, for the physicist who assumes a realist stance the question of probability is {\it necessarily} related to experience and physical reality.

Luigi Accardi proved in 1981 that there is a direct relation between Bell inequalities and probability models \cite{Accardi82}. The theorem of Accardi states that any theory which violates Bell inequality has a non-Kolmogorovian probability model. Since only Kolmogorovian models can be interpreted in terms of referring to a degree of ignorance of a presupposed state of affairs given by a set of definite valued preexistent properties, this means that QM possesses a probability model which cannot be interpreted in terms of ignorance of such preexistent reality. The fact that QM possesses a non-Kolmogorovian probability model is not such a big issue from a mathematical perspective: many mathematicians work with these probability structures and do not get astonished in any way by them. But from a realistic representational physical perspective, the question which arises is very deep, namely, what is the meaning of a concept of probability which does not talk about the degree of knowledge of a definite state of affairs? From our perspective, if such a question is not properly acknowledged, the statement ``QM is a theory about probabilities'' looses all physical content. It might be regarded as either an obvious mathematical statement with no interest ---it only states the well known fact that in QM there is a (non-Kolmogrovian) probability measure assigned via Gleason's theorem--- or a meaningless physical statement, since we do not know what quantum probability is in terms of a physical concept. According to our stance, if we are to understand QM as a physical theory, and not merely as a mathematical or algorithmic structure, it is clear that we still need to provide a link between the mathematical structure and a set of physical concepts which are capable of providing a coherent account of quantum phenomena.

\section{Empirical Terms vs Physical Concepts}

Logical positivists fought strongly against dogmatic metaphysical thought, imposing a reconsideration of observability beyond the {\it a priori} categories of Kantian metaphysics. In their famous {\it Manifesto} \cite{VC} they argued that: ``Everything is accessible to man; and man is the measure of all  things. Here is an affinity with the Sophists, not with the Platonists; with the Epicureans, not with the Pythagoreans; with all those who stand for earthly being and the here and now.'' Their main attack to metaphysics was designed through the idea that one should focus in ``statements as they are made by empirical science; their meaning can be determined by logical analysis or, more precisely, through reduction to the simplest statements about the empirically given.'' Their architectonic stood on the distinction between {\it empirical terms}, the empirically ``given'' in physical theories, and {\it theoretical terms}, their translation into simple statements. This separation and correspondence between theoretical statements and empirical observation would have deep consequences not only regarding the problems addressed in philosophy of science but also with respect to the limits of development of many different lines of research. The important point is that even though within the philosophy of science community this distinction has been strongly criticized and even characterized as ``naive''; many of the problems discussed in the literature still presuppose it implicitly. Indeed, as remarked by Curd and Cover: 

\begin{quotation}
\noindent {\small``Logical positivism is dead and logical empiricism is no longer an avowed school of philosophical thought. But despite our historical and philosophical distance from logical positivism and empiricism, their influence can be felt. An important part of their legacy is observational-theoretical distinction itself, which continues to play a central role in debates about scientific realism.''  \cite[p. 1228]{PS}}\end{quotation}

\noindent One of the major consequences of this ``naive'' perspective towards observation is that physical concepts become supplementary elements in the analysis of physical theories. Indeed, when a physical phenomenon is understood as independent of physical concepts and metaphysical presuppositions, {\it empirical terms} configure an objective set of data which can be directly related ---without any metaphysical constrain--- to a formal scheme. Actual empirical observations become then the very fundament of physical theories which, following Mach, should be understood as providing an ``economical'' account of such observational data. As a consequence, metaphysics and physical concepts are completely out of the main picture.

\bigskip

\begin{center}
{\it Empirical Data ---------------  Theoretical Terms \\
\bigskip
(Supplementary Interpretation)}
\end{center}

\bigskip

According to this scheme, physical concepts are not essentially needed since the analysis of a theory can be done by addressing the logical structure which accounts for the empirical data. The role of concepts becomes then accessory: adding metaphysics might help us to picture what is going on according to a theory. Like van Fraassen argues \cite{VF10}, it might be interesting to know what the world is like according to an interpretation of a formalism. However, one can perfectly do without interpretation when the question addressed is only related to empirical findings. Many realists within philosophy of physics, while they stress the need of an interpretation accept the idea that the formalism already provides direct access to empirical data. Like a Troyan horse, these realist schemes have hidden within the main weapon (metaphysical presupposition) of the enemy (empiricism). Indeed, in philosophy of physics both realists and empiricists seem to agree that metaphysical schemes are only necessary when attempting to ``understand'' ---a term which remains dependent on the philosophical stance--- a physical theory. Their distance seems to be the strength with which they argue for or against the need of interpretation. This is the main reason why the ``interpretation'' of a theory has been understood in philosophy of physics as something ``added'' to an already formalized empirical theory. Thus, physical concepts are not directly related to the metaphysical foundation of phenomena and experience.\footnote{A clear expression of this situation is the so called ``underdetermination problem'' which implicitly assumes that a theory can account for phenomena independently of a metaphysical scheme.} 

Against this fundamentally empiricist based perspective ---extensively widespread within philosophy of physics even in the context of supposedly realist perspectives--- we understand that each physical theory is a triad composed by a {\it mathematical formalism}, a {\it conceptual network} and a limited, specific {\it field of phenomena}. In this scheme physical notions play a fundamental role. {\it Physical concepts are defined through metaphysical principles which configure and determine physical experience itself.} Physical observation cannot be considered in terms of ``common sense'' realism, {\it physical observation is always theory-laden}. We are always have to deal with a physical (and metaphysical) representation; and thus, the analysis must always begin, not by collecting a set of ``naked'' empirical data, but by considering and making explicit the metaphysical presuppositions related to the theory and its phenomena. Our philosophical post-Kantian realist position stresses the need to consider physical notions as fundamental elements of a theory, without which physical observation of phenomena cannot be defined. As obvious as it is, a `field' cannot be observed without the notion of field, we simply cannot observe a `particle' or a `wave' without presupposing such physical concepts. These concepts are undoubtedly part of a metaphysical architectonic developed through centuries. Naturalizing such concepts in terms of ``common sense'' {\it givens} is turning a specific metaphysical scheme into dogma.

\begin{quotation}
\noindent {\small``Concepts that have proven useful in ordering things easily
achieve such an authority over us that we forget their earthly
origins and accept them as unalterable givens. Thus they come to be
stamped as `necessities of thought,' `a priori givens,' etc. The
path of scientific advance is often made impossible for a long time
through such errors.'' \cite[p. 102]{Einstein16}}
\end{quotation}

In particular, it is important to remark that all classical physical entities ---as we discussed above--- presuppose the metaphysical PE, PNC and PI. These are not principles that are found or observed in the world, but the very conditions that allow us to determine physical experience itself \cite{deRonde14}. As remarked by Einstein in his famous recommendation, which led Heisenberg to the principle of indetermination: ``It is only the theory which can tell you what can be observed." Einstein's philosophical position has been many times carichaturized in the literature as a scientific realist.\footnote{As remarked by Howard \cite[p. 73]{Howard05}, Einstein was certainly part of the neo-Kantian tradition: ``Einstein was dismayed by the Vienna Circle's ever more stridently anti-metaphysical doctrine. The group dismissed as metaphysical any element of theory whose connection to experience could not be demonstrated clearly enough. But Einstein's disagreement with the Vienna Circle went deeper. It involved fundamental questions about the empirical interpretation and testing of theories."} The fact that ``he was not the friend of any simple realism" \cite[p. 206]{Howard93} can be witnessed from the very interesting remark, recalled by Heisenberg, in which Einstein explained: 

\begin{quotation}
\noindent {\small ``I have no wish to appear as an advocate of a
naive form of realism; I know that these are very difficult
questions, but then I consider Mach's concept of observation also
much too naive. He pretends that we know perfectly well what the
word `observe' means, and thinks this exempts him from having to
discriminate between `objective' and `subjective' phenomena. No
wonder his principle has so suspiciously commercial a name: `thought
economy.' His idea of simplicity is much too subjective for me. In
reality, the simplicity of natural laws is an objective fact as
well, and the correct conceptual scheme must balance the subjective
side of this simplicity with the objective. But that is a very
difficult task." A. Einstein quoted by W. Heisenberg
\cite[p. 66]{Heis69}}\end{quotation}

\noindent Einstein's position was orthodoxy at the time. Most of the founding fathers of QM ---exception made of Dirac--- were also part of this same neo-Kantian tradition which understood that the observation of physical phenomena was metaphysically constrained. Following Einstein's dictum, Heisenberg went also against the positivist interpretation of empirical science as disconnected from metaphysical presuppositions:  

\begin{quotation}
\noindent {\small ``The history of physics is not only a sequence of experimental discoveries and observations, followed by their mathematical description; it is also a history of concepts. For an understanding of the phenomena the first condition is the introduction of adequate concepts. Only with the help of correct concepts can we really know what has been observed."  \cite[p. 264]{Heis73}}\end{quotation}

Going back to our scheme, the three elements that compose a physical theory form a perfect circle with no preeminence of one over the other. All three elements are interrelated in such a way that only through their mutual inter-definition we can access physical experience. 

\bigskip

\begin{center}
{\it Conceptual Network   \ \ \ \ \ \ \ \ \ \ \ \ \ \  \ \ \    Mathematical Formalism \\
\bigskip
\bigskip
Field of Phenomena}
\end{center}

\bigskip

\noindent In order to be clear about our perspective of analysis we would like to make explicit our philosophical stance which takes into account metaphysical representation and the theory ladenness of physical observation right from the start:\\ 

\noindent {\it {\bf Representational Realist Stance (RRS):} A representational realist account of a physical theory must be capable of providing a physical (and metaphysical) representation of reality in terms of a network of concepts which coherently relates to the mathematical formalism of the theory and allows to make predictions of a definite field of phenomena (expressed through such concepts).}\\

\noindent According to our RRS physical statements about phenomena must be necessarily related to the physical (and metaphysical) representation provided by the theory in terms of definite physical concepts. There is no ``common sense'' experience in physics. Every experience in physics is a restricted experience, constrained by physical concepts and metaphysical presuppositions. Physics is in essence a metaphysical enterprise. As Einstein remarked: ``The problem is that physics is a kind of metaphysics; physics describes `reality'. But we do not know what `reality' is. We know it only through physical description...'' From this perspective, the problem with the orthodox formalism of QM is that it provides predictions which have not been coherently related to a network of adequate physical concepts. To say it shortly, we do not know what QM is talking about. But if we do not know how to account for the `clicks' in detectors in a typical quantum experiment, we simply don't understand either what is a quantum phenomenon. Still today, the `quantum clicks' that we find in the lab do not have a conceptual or metaphysical support. Boole-Bell type inequalities have proven that `quantum clicks' lie outside the scope of classical local-realistic theories.\footnote{As remarked by Itamar Pitowsky \cite[p. 95]{Pitowsky94}: ``In the mid-nineteenth century George Boole formulated his `conditions of possible experience'. These are equations and inequalities that the relative frequencies of (logically connected) events must satisfy. Some of Boole's conditions have been rediscovered in more recent years by physicists, including Bell inequalities, Clauser Horne inequalities, and many others.''} What we do know in fact is that, if we respect the orthodox formalism of QM, those `clicks' are not being produced by classical entities.

\section{The EPR-Battle: Counterfactual Statements and (Actual) Elements of Physical Reality}

The power of physics comes from its amazing predictive capacity; something that is exposed through the empirical confirmation of (operational) counterfactual statements. If a theory is empirically adequate then counterfactual statements of the type: ``if we measure physical quantity $A$, the result will be $x$; but if we measure instead physical quantity $B$ the result will be $y$'' are always considered to be what the theory is talking about. As clearly expressed by Griffiths \cite[p. 361]{Griffiths02}: ``If a theory makes a certain amount of sense and gives predictions which agree reasonably well with experimental or observational results, scientists are inclined to believe that its logical and mathematical structure reflects the structure of the real world in some way, even if philosophers will remain permanently skeptical." Indeed, operational counterfactual statements conform the core of the objective physical reality the theory describes. Counterfactual reasoning is a {\it necessary condition} not only for constructing a representation that provides an objective account of physical reality independent of the choices and actions of subjects but also for physical discursivity itself. We should remark, due to the ongoing debate about counterfactuals in QM, that these kind of physical statements need not be necessarily related to ``possible worlds'' or to ``the reification of modalities'' ---a particular way of analyzing these subjects by logicians and analytic metaphysics which has also penetrated deeply philosophy of physics. In physics, operational counterfactual reasoning does not imply that every statement is actually real. Obviously, the fact that I can imagine an experience in the future does not imply its reality. Operational counterfactual reasoning has been assumed in every physical theory that we know and allows a theory to make predictions in terms of meaningful physical statements. \\

\noindent {\it {\bf Meaningful Physical Statements (MPS):} If given a specific situation a theory is capable of predicting in terms of definite physical statements the outcomes of possible measurements, then such physical statements are meaningful relative to the theory and must be considered as constitutive parts of the particular representation of physical reality that the theory provides. Measurement outcomes must be understood only as exposing the empirical adequacy (or not) of the theory.}\\

\noindent MPS are not necessarily statements about future events, such as for example ``if I measure the spin in the $x$-direction, I will obtain spin-up with probability 0.4 and spin-down with probability 0.6." MPS can be also statements about the past or the present. For example, according to some physical theories I can claim that ``the earth was formed about 4.54 billion years ago'' (long before even physics was imagined!), or that if someone would perform a free fall experiment in the moon at this very moment, due to its gravity, ``the object would be falling accelerated at 1.6 $\frac{m}{s^2}$." That is indeed the magic of both physics (and metaphysics), the possibility to represent, think and imagine beyond the here and now.

Physical statements that allow to predict certain phenomena have been always intuitively related to physical reality. According to physicists, if we possess an empirically adequate physical description of a state of affairs we can predict what will happen in any particular experiment.\footnote{It is interesting to notice that in such kind of statements we see the two main understandings of actuality coming together: the actuality {\it hic et nunc} of observations is an expression of the actual {\it preexistent} mode of existence of properties. Unfortunately, it is very frequent to find in the literature a mixture between these two different meanings of actuality.} For instance, we also know what might have happened if I had performed an experiment in the past or in the present, in a different place to the one I am now. Experiments in classical physics allow us to learn about the preexistent properties of a system. The strong realist presupposition is that once we have an empirically adequate theory we don't even need to perform an experiment in order to know the result! Take for example a physical object as a small ball, one can imagine all the possible experiments that one could perform inside a lab with it. We know the acceleration of a ball in a free fall experiment on earth will be 9.8  $\frac{m}{s^2}$ and we can also predict the motion of the ball if we through it inside the room. There are indeed many experiments we could perform of which we know the answer beforehand by simply calculating their results using classical mechanics. There is no single physicist that would dare go against the predictions of classical Newtonian mechanics. And that is the whole beauty of physics, at least from a realist perspective: physical representations talk about physical reality independently of the here and now. 

The importance of counterfactual statements as related to physical reality was stressed, in the context of QM, by Einstein himself in 1935, in a famous article which would be known as the ``EPR paper'' \cite{EPR}. In it, Einstein together with his students Podolsky and Rosen, used his famous definition of what was to be considered an {\it element of physical reality} in order to show that QM seemed not to be a {\it complete theory}. As remarked by them: ``Whatever the meaning assigned to the term {\it complete}, the following requirement for a complete theory seems to be a necessary one: {\it every element of physical reality must have a counterpart in the physical theory.}'' [{\it Op. cit.}, p. 777] Indeed, this seems to be a necessary condition for a theory which attempts to provide an account of physical reality. However, Einstein's definition stressed only a limited set of the MPS predicted by quantum theory. His definition focused only on those MPS which could be related to an actualist metaphysical account of physical reality ---leaving aside the more general probabilistic statements.\\  

\noindent {\it {\bf Einstein's (Actual) Element of Physical Reality:} If, without in any way disturbing a system, we can predict with certainty (i.e., with probability equal to unity) the value of a physical quantity, then there exists an element of reality corresponding to that quantity.}\\

\noindent As remarked by Aerts and Sassoli \cite[p. 20]{AertsSassoli}: ``the notion of `element of reality' is exactly what was meant by Einstein, Podolsky and Rosen, in their famous 1935 article. An element of reality is a state of prediction: a property of an entity that we know is actual, in the sense that, should we decide to observe it (i.e., to test its actuality), the outcome of the observation would be certainly successful.'' Indeed, certainty and actuality were the restrictive constraints of what could be considered in terms of physical reality. 

But Bohr, contrary to Einstein, had a very different standpoint regarding the meaning of physics in general, and of QM in particular. In his 1935 reply paper to EPR \cite{Bohr35} ---appeared in the following volume of {\it Physical Review}--- he argued that in QM things were completely different to any other physical theory. Bohr wanted to presuppose classical discourse in terms of classical notions (e. g., waves and particles) even at the price of restricting the multiple contexts of analysis provided by the formalism itself. Even though each basis was directly related to the correct predictions of statistical outcomes of observables, it was argued that in order to discuss about quantum properties the very precondition was the choice of a single context (interpreted in terms of an experimental arrangement). In this way, quantum physics had been restricted to the here and now experimental set-up. At the same time, Bohr  \cite[p. 7]{WZ} had strongly argued about the impossibility of providing a physical representation of QM beyond classical notions, claiming that: ``[...] the unambiguous interpretation  of any measurement must be essentially framed in terms of classical physical theories, and we may say that in this sense the language of Newton and Maxwell will remain the language of physicists for all time.'' According to him [{\it Op. cit.}, p. 7], ``it would be a misconception to believe that the difficulties of the atomic theory may be evaded by eventually replacing the concepts of classical physics by new conceptual forms.'' The choice of Bohr was to stick to classical discourse and give up operational counterfactual reasoning of MPS in QM ---which was explicitly used within the EPR argument. Bohr was willing to develop a new complementarity scheme even at the price of abandoning the physical representation of quantum reality. 

Bohr took as a standpoint the idea that observed measurement outcomes were perfectly defined in QM and added the necessity of choosing a particular context between the many possible ones\footnote{A problem known today in the literature as the infamous basis problem. One that has found no true solution until the present.} in order to recover a classical ``{\it as-if} discourse'' in terms of ``waves'' and ``particles''. 

\begin{quotation}
\noindent {\small``[...] the choice between the experimental procedures suited for the prediction of the position or the momentum of a single particle which has passed through a slit in a diaphragm, we are, in the `freedom of choice' offered by the last arrangement, just concerned with the {\it discrimination between different experimental procedures which allow of the unambiguous use of complementarity classical concepts.}''  \cite[p. 699]{Bohr35}}
\end{quotation}

\noindent But this complementarity scheme designed by the Danish physicist precluded ---since it denied operational counterfactual reasoning itself--- the very possibility of relating MPS to an objective physical description of reality ---independent of {\it subjective choices}. After Bohr's reply  \cite{Bohr35}, unlike a classical object which preexists (in terms of definite valued properties) independently of the choice of any experiment, it was accepted that quantum systems and properties were explicitly dependent on the choice of an experimental arrangement or context. 

Once and again it was repeated that Bohr had been ``the true winner of the EPR battle'' ---as well as of the Solvay confrontation some years before. However, no one could really explain why. Bohr had designed a contradictory algorithmic language based on his complementarity principle according to which, it only made sense to talk about ``waves'' and ``particles'' once the choice of an experimental set up had been performed by the physicist in the lab. Subjectivity had been introduced for the first time {\it within} physical description, creating what is known today as ``the quantum omelette'' (see for discussion \cite{deRonde15d}). As most clearly stated by Jaynes: 

\begin{quotation}
\noindent {\small``[O]ur present [quantum mechanical] formalism is not purely epistemological; it is a peculiar mixture describing in part realities of Nature, in part incomplete human information about Nature ---all scrambled up by Heisenberg and Bohr into an omelette that nobody has seen how to unscramble. Yet we think that the unscrambling is a prerequisite for any further advance in basic physical theory. For, if we cannot separate the subjective and objective aspects of the formalism, we cannot know what we are talking about; it is just that simple.''  \cite[p. 381]{Jaynes}}
\end{quotation}

But to be fare, it was not true that Bohr wanted to discuss about reality. His scheme was totally consistent and very difficult to tackle. It rested on an understanding of physics in highly pragmatic terms, as a ``tool'' to approach intersubjective agreement between experimentalists. In this respect, Bohr was indeed much closer to logical positivism than he himself would have admitted. 

\begin{quotation}
\noindent {\small ``Physics is to be regarded not so much as the study of something {\it a priori} given, but rather as the development of methods of ordering and surveying human experience. In this respect our task must be to account for such experience in a manner independent of individual subjective judgement and therefor objective in the sense that it can be unambiguously communicated in ordinary human language." 
\cite{Bohr60}} \end{quotation}

\noindent Even though Bohr was a neo-Kantian himself which understood phenomena as related to metaphysical principles, against metaphysical questions and problems, he was in line with the positivist appeal to Sophists and their epistemological perspective which placed the subject ---and his here and now experience--- as the fundament of knowledge itself. Bohr, contrary to the logical positivists who presupposed a ``common sense'' here and now observation, had placed his {\it a priori} in classical language ---which had the purpose of constraining phenomena as classical space-time phenomena.  Bohr had reintroduced Protagoras {\it dictum} within physics, adding to it the importance of classical language and connecting physics to the main philosophical debate of the 20th century: the linguistic turn. His new precept could be read in the following terms: {\it the subject and his (classical) language are the measure of al things}. Bohr was not interested in the problem of reality. Instead of getting into the riddle of an ontological analysis Bohr focused on epistemological concerns. Indeed, as remarked by A. Petersen, his long time assistant: 

\begin{quotation}
\noindent {\small ``Traditional philosophy has accustomed us to regard
language as something secondary and reality as something primary.
Bohr considered this attitude toward the relation between language
and reality inappropriate. When one said to him that it cannot be
language which is fundamental, but that it must be reality which, so
to speak, lies beneath language, and of which language is a picture,
he would reply, ``We are suspended in language in such a way that we
cannot say what is up and what is down. The word `reality' is also a
word, a word which we must learn to use correctly.'' Bohr was not
puzzled by ontological problems or by questions as to how concepts
are related to reality. Such questions seemed sterile to him. He saw
the problem of knowledge in a different light." \cite[p. 11]{Petersen63}}
\end{quotation}

\noindent Bohr wanted to develop a language that would allow us to account for phenomena in terms of classical physical concepts, even at the price of dissolving the relation between such concepts and reality. His complementarity approach was designed in order to support the inconsistencies of such incompatible relations. This had very important consequences for the development of physics. As Fine makes the point: 

\begin{quotation}
\noindent {\small ``[The] instrumentalist moves, away from a realist construal
of the emerging quantum theory, were given particular force by
Bohr's so-called `philosophy of complementarity'; and this
nonrealist position was consolidated at the time of the famous
Solvay conference, in October of 1927, and is firmly in place today.
Such quantum nonrealism is part of what every graduate physicist
learns and practices. It is the conceptual backdrop to all the
brilliant success in atomic, nuclear, and particle physics over the
past fifty years. Physicists have learned to think about their
theory in a highly nonrealist way, and doing just that has brought
about the most marvelous predictive success in the history of
science.'' \cite[p. 1195]{PS}}
\end{quotation}

\noindent Contrary to Fine we do not understand this as ``the most marvelous'' epoch of science but rather as a quite obscure period in which we have not advanced much in really understanding one of the main theories of the 20th Century. After more than one century after its creation we still don't know what QM is talking about. Contrary to Bohr, we believe it is possible to develop a (non-naive) post-Katian realist position, one that understands that every physical theory must presupposed objective physical representation of reality, and that in this respect QM cannot be an exception. Giving up physical representation would be giving up physics itself, for the lack of representation transforms physics into mere technique. 

Our RRS attempts to bring back metaphysical considerations in the analysis of QM by taking into account three main desiderata: the first is that physical observation is theory-laden and thus always metaphysically founded; the second is that operational counterfactual reasoning about MPS is the kernel of physical discourse and in consequence cannot be abandoned if we seek to find an objective representation of physical reality; the third and final desideratum is that predictions must be necessarily related to the physical representation of reality provided by the theory. Against Bohr, (actual) experiments and measurements cannot be regarded as the point of departure, since it is only the theory which can tell you what can be observed. We need to do exactly the opposite, we need to read out an objective physical description from the formalism escaping at the same time dogmatic classical metaphysics ---which is today still grounded in the metaphysical notion of actuality. Just like Einstein taught us to do in Relativity, we need to concentrate on what the theory predicts, and be ready to come up with new physical concepts that match the formalism and explain phenomena. From our perspective ---contrary to Bohr's dogmatism with respect to classical language and physical experience---, every new physical theory determines a radically new field of experience which is necessarily related to a language constituted by new physical concepts.

\section{Actual Properties and Observation in QM}

The general metaphysical principle implied by the understanding of Newtonian mechanics, that `Actuality = Reality', has become an unquestionable dogma within physics. As a silent fundament all of physics has been developed following the metaphysics of actuality. And even though QM was born from a deep positivist deconstruction of the {\it a priori} classical Newtonian notions ---and in this sense the philosophy of Mach can be understood as the very precondition for the creation of both QM and relativity theory--- it was very soon reestablished within the limits of classical metaphysics itself. The constrains of actuality have been unquestionably accepted by philosophers of physics either in terms of {\it hic et nunc} observation (empiricism and its variants) or as the mode of {\it preexistence} of properties (realism). Both positions have remained captive of actualism; trapped in the metaphysical net designed (through the PE, PNC and PI) by Aristotle around the 5th century before Christ and imposed by Newton in the 18th Century of our time. Actual (preexistent) properties and actual (here and now) observations are two sides of the same (metaphysical) coin.

Today, both realists and anti-realists support an anti-metaphysical understanding of observation within philosophy of physics. It is accepted by both parties that QM is an empirically adequate theory and, consequently, that the problem is not related to the understanding of quantum phenomena. This can be directly linked not only to the positivist ``common sense'' or ``naive'' understanding of observation, but also ---maybe more importantly--- to Bohr's analysis of QM based on the idea that any physical phenomena is a classical phenomena, or in his own words, to the idea that ``[...] the unambiguous interpretation  of any measurement must be essentially framed in terms of classical physical theories''. In the present, van Fraassen \cite[pp. 202-203]{VF80} has followed this path developing an empiricist stance according to which: ``To develop an empiricist account of science is to depict it as involving a search for truth only about the empirical world, about what is actual and observable.'' Making explicit the fact that observation should be understood in terms of ``common sense'' observation. The problem, according to van Fraassen, only appears with respect to the ``non-observable'' entities ---such as e.g, an atom or an electron. Instrumentalists assume exactly the same ground (as empiricist) considering actual observation in terms of ``common sense'' observation of measurement outcomes. As made explicit by Fuchs and Peres: ``[...] quantum theory does not describe physical reality. What it does is provide an algorithm for computing probabilities for the macroscopic events (``detector clicks") that are the consequences of experimental interventions." On the other hand, realists approaches to QM have focused on Einstein's implicit use of the {\it elements of physical reality} in terms of actuality and his recommendation to extend QM ---which should be considered as ``incomplete''--- to a more general formalism, one that goes back to a description in terms of actual properties  restoring ``a classical way of thinking about {\it what there is}.'' \cite{Bacciagaluppi96}  This idea, presupposed by the Hidden Variable Program (HVP), has also permeated strongly the rest of realist interpretations of QM, which in one way or the other have ended up always discussing in terms of actual properties, grounded as well on ``common sense'' observation. Such is the case of the modal interpretation of Dieks, Griffiths' consistent histories approach, and the many worlds interpretation. Even those interpretations such as the ones proposed by Heisenberg, Popper, Margenau and Piron, that have argued in favor of considering a different realm to that of actuality, were not able to advance in an ontological definition of such realm. Potentialities, propensities and dispositions have been repeatedly defined only in terms of {\it a process of becoming actual} (see for discussion \cite{deRonde11}). These teleological schemes have betrayed, because of their standpoint and focus on the measurement problem, any true possibility of progress and development beyond the actual realm. 

Actuality imposes a mode of being (of both properties and observations) determined by the PE, the PNC and the PI. Everything is reduced then either to: {\it yes-no properties} or {\it yes-no experimental observations}. But what if QM cannot be subsumed under the metaphysical equation imposed by Newtonian physics: Actuality = Reality?

\section{Revisiting Quantum Physical Reality}

We believe it would be no exaggeration to claim that the EPR paper together with Bohr's reply, have determined the fate of QM up to the present. The EPR paper ended with a recommendation to extend QM in order to recover a classical actualist understanding about what there is: 

\begin{quotation}
\noindent {\small``While we thus have shown that the wave function does not provide a complete description of the [actual] physical reality, we left open the question of whether or not such a description [of actual properties] exists. We believe, however, that such theory is possible.''  \cite[p. 780]{EPR}}
\end{quotation}

\noindent Bohr argued instead that: 

\begin{quotation}
\noindent {\small``While [...] in classical physics the distinction between object and measuring agencies does not entail any difference in the character of the description of the phenomena concerned, it fundamental importance in quantum theory [...] has its root in the indispensable use of classical concepts in the interpretation of all proper measurements, even tough the classical theories do not suffice in accounting for the new types of regularities with which we are concerned in atomic physics.''  \cite[p. 701]{Bohr35}}
\end{quotation}

EPR was the final battle of the two main figures in the physics of the 20th century, and even though Bohr was declared the only triumphant survivor, both lines were developed under the constraints and limits of the logical positivist ``naive'' or ``common sense'' understanding of observation ---abandoning one of the main discussions of the founding fathers of QM regarding the meaning of quantum phenomena and observation. After the clash of the two titans, physicists were confronted with a choice between two different paths. Either they could follow Bohr and be satisfied with an inconsistent intersubjective language ruled by complementarity (see \cite{KrausedaCosta}) with no direct reference to physical reality, or they could follow Einstein and try to find a new formalism that would allow them to recover a classical actualist type-description of physical (quantum) reality. But this crossroad, imposed by Einstein, Bohr and logical positivism, hides a road sign called {\it Wolfgang Pauli} which exposed the lines of a more radical resolution to the quantum riddle. Indeed, between the founding fathers, we regard Wolfgang Pauli as the most radical and revolutionary thinker of all. Against Bohr, he stood always close to metaphysics and the problem of reality; beyond Einstein, he was ready to reconsider the meaning and definition of physical reality itself.\footnote{Which for Einstein was determined through space-time separability.} 

\begin{quotation}
\noindent {\small ``When the layman says `reality' he usually thinks that he is speaking about something which is self-evidently known; while to me it appears to be specifically the most important and extremely difficult task of our time to work on the elaboration of a new idea of reality.'' \cite[p. 193]{Laurikainen98}}
\end{quotation}

What do we mean when we say that ``particles are physically real according to classical mechanics''? Following Heisenberg and his closed theory approach, this question has a definite answer according to our RRS. It means that classical mechanics is capable of expressing through the relation between the formalism of mathematical calculus and a network of concepts ---such as, for example, space, time, particle, mass, position, velocity--- a broad field of phenomena. It is in this sense that the notion of particle is a metaphysical machinery which allows us to express reality. In QM we have a sound formalism, with features such as contextuality, superposition and indetermination which defy a realist classical scheme in terms of an ASA. However, all approaches until today have stood close to physical reality understood in an actualist fashion. Because of this, the features of the quantum formalism have been regarded as ``problems'' which we need to bypass or overcome in order to recover our classical way of thinking about {\it what there is}. Our proposed line of research, following Pauli, is to turn the problem upside-down. What we need to do is to develop physical reality according to what QM needs it to be. 

Heisenberg's closed theory approach is the key to abandon another presupposed dogma ---also imposed by Bohr--- according to which QM must be related to classical physics in terms of a {\it limit}. Once we accept the possibility of considering an independent metaphysical scheme to account for QM, the ``problems'' addressed in the literature become instead essential features of the metaphysical system we need to construct in order to coherently relate the formalism with physical reality and experience. We need to develop a new way of understanding reality beyond the ruling of actuality.  To escape the ruling of actuality ---both in terms of {\it hic et nunc} observation and {\it preexistent} properties--- means to abandon, on the one  hand, the idea that we have a clear definition of what is observed according to QM, and on the other hand, the idea that actuality is the only possible way to conceive and understand physical reality. Our strategy is to take as a standpoint the formalism and its predictive power in order to develop new physical concepts which relate coherently to the formalism and can allow us to think about the physical meaning of quantum phenomena.

\section{Objective Probability and Generalized Elements of Physical Reality}

The quantum wave function $\Psi$ provides definite physical statements regarding observables through the Born rule. The MPS provided by QM, statements that have been used in order to develop experimental situations and outstanding technological developments, are of the following type:  

\begin{definition}
{\bf MPS in QM:} Given a vector in Hilbert space, $\Psi$, the Born rule allows us to predict the average value of (any) observable $O$. 
$$\langle \Psi| O | \Psi \rangle = \langle O  \rangle$$
This prediction is independent of the choice of any particular basis. 
\end{definition}

To take seriously the QM formalism means for us to take into account all the predictions provided by QM; i.e., both {\it certain} (probability equal to unity) and {\it statistical} (probability between zero and unity) predictions about physical quantities. We need to create a new understanding of probability in terms of objective knowledge abandoning its classical understanding in terms of ignorance about an actual state of affairs. But how to do so in relation to physical reality? We believe that a good standpoint is the generalization of Einstein's realist definition of an element of physical reality. The redefinition must keep the relation imposed between predictive statements and reality, but leave aside both the actualist constraint imposed by certainty (probability equal to unity) and the importance of measurement which should be only regarded as confirming or disconfirming a specific prediction of a theory.\\

\noindent {\it {\bf Generalized Element of Physical Reality:} If we can predict in any way (i.e., both probabilistically or with certainty) the value of a physical quantity, then there exists an element of reality corresponding to that quantity.}\\

By extending the limits of what can be considered as physically real, we have also opened the door to a new understanding of QM beyond classical physics. The problem is now set: we now need to find {\it a physical concept that is capable of being statistically defined in objective terms.} That means to find a notion that is not defined in terms of yes-no experiments (as it is the case of classical properties), but is defined instead in terms of a probabilistic measure. Of course, this first step must be accompanied by developing a network of physical notions that accounts for what QM is talking about, beyond measurement outcomes. In the end, our new non-classical physical scheme will also have to be capable of taking into account the main features brought in by the orthodox formalism. 

\begin{enumerate}

\item[i.]  {\it Our network of physical concepts must provide a deeper understanding of the principle of indetermination, the principle of superposition, the quantum postulate and quantum phenomena in general.} 

\item[ii.]  {\it Our network of physical concepts must also explain the physical meaning of non-locality, non-separability and quantum contextuality.} 

\item[iii.]  {\it Our metaphysical scheme must be capable of recovering an objective notion of measurement (one that exposes a preexistent state of affairs).} 

\item[iv.]  {\it The physical representation must account for all MPS in QM respecting (operational) counterfactual reasoning (as in any physical theory).} 

\end{enumerate}

\noindent We believe it is possible to come up with a physical network of concepts that takes into account these features. The price to pay is to abandon the metaphysics of actuality and construct a new non-classical metaphysical scheme with physical concepts specifically designed in order to account for the orthodox formalism of QM.

\section{Potential Powers in QM}

Our research has analyzed the idea of considering a mode of existence truly independent of actuality, namely, ontological potentiality. Elsewhere, we have introduced ontological potentiality as the realm of which QM talks about. This realm is defined by the principles of indetermination, superposition and difference. In order to advance we also need to introduce the notions of {\it Potential State of Affairs} (PSA), {\it potential effectuation} and {\it immanent cause} \cite{deRonde13, deRonde15a}. Indeed, by developing these new concepts, we expect that our formal analysis regarding {\it quantum possibility} \cite{RFD14} can find a coherent physical interpretation. But now the question arises: what are the ``things'' which exist and interact within this potential realm? Our answer is: {\it powers} with definite {\it potentia}.  Indeed, while entities are composed by properties which exist in the actual mode of being, we have argued that an interesting candidate to consider what exists according to QM is the notion of {\it power}. Elsewhere \cite{deRonde11, deRonde13}, we have put forward such an ontological interpretation of powers. In the following we summarize such ideas and provide an axiomatic characterization of QM in line with these concepts. 

{\it The mode of being of a power is potentiality}, not thought in terms of classical possibility (which relies on actuality) but rather as a mode of existence ---i.e., in terms of ontological potentiality. To possess the power of {\it raising my hand}, does not mean that in the future `I {\it will} raise my hand' {\it or} that in the future `I {\it will not} raise my hand'; what it means is that, here and now, I possess a power which exists in the mode of being of potentiality, {\it independently of what happens or will happen in actuality}. Powers do not exist in the mode of being of actuality, they are not actual existents, they are undetermined potential existents. Powers, like classical properties, preexist to observation, but unlike them preexistence is not defined in the actual mode of being as an Actual State of Affairs (ASA), instead we have a {\it potential preexistence} of powers which determines a Potential State of Affairs (PSA). While an ASA can be defined in terms of a set of actual properties, a PSA is defined as a set of powers with definite potentia. {\it Powers are indetermined.} The concept of `power' allow us to compress experience into a picture of the (quantum) world, just like entities such as particles, waves and fields, allow us to do so in classical physics. We cannot ``see" powers in the same way we see objects.\footnote{It is important to notice there is no difference in this point with the case of entities: we cannot ``see" entities ---not in the sense of having a complete access to them. We only see perspectives which are unified through the notion of object.} Powers are experienced in actuality through {\it elementary processess}. A power is sustained by a logic of actions which do not necessarily take place, it \emph{is} and \emph{is not}, {\it hic et nunc}. 

A basic question which we have posed to ourselves regards the ontological meaning of a {\it quantum superposition} \cite{deRonde15c}. What does it mean to have a mathematical expression such as: $\alpha | \uparrow \rangle + \beta  | \downarrow \rangle$, which allows us to predict precisely, according to the Born rule, experimental outcomes? We believe that quantum superpositions are the central elements of QM. But what is the physical representation of these quantum superpositions? Our theory of powers has been explicitly developed in order to try to find an answer to this particular question. Given a superposition in a  particular basis, $\Sigma \  c_i | \alpha_i \rangle$, the powers are represented by the elements of the basis, $| \alpha_i \rangle$, while the coordinates in square modulus, $|c_i|^2$, are interpreted as the potentia of each respective power. {\it Powers can be superposed to different ---even contradictory--- powers} \cite{deRonde15a}. We understand a quantum superposition as encoding a set of powers each of which possesses a definite {\it potentia}. This we call a {\it Quantum Situation} ($QS$). For example, the quantum situation represented by the superposition $\alpha | \uparrow \rangle + \beta |\downarrow\rangle$, combines the contradictory powers, $| \uparrow \rangle$ and $|\downarrow\rangle$, with their respective potentia, $|\alpha|^2$ and $|\beta|^2$. Contrary to the orthodox interpretation of the quantum state, we do not assume the metaphysical identity of the multiple mathematical representations given by different bases \cite{deRonde15c}. Each superposition is basis dependent and must be considered as a distinct quantum situation. For example, the superpositions $c_{x1} | \uparrow_{x} \rangle + c_{x2} |\downarrow_{x}\rangle$ and $c_{y1} | \uparrow_{y} \rangle + c_{y2} |\downarrow_{y}\rangle$, which are representations of the same $\Psi$ and can be derived from one another via a change in basis, are interpreted as two different quantum situations, $QS_{\Psi, B_x}$ and $QS_{\Psi, B_y}$. 

The logical structure of a superposition is such that a power and its opposite can exist at one and the same time, violating the principle of non-contradiction \cite{daCostadeRonde13}. Within the faculty of raising my hand, both powers (i.e., the power `I am {\it able to} raise my hand' and the power `I am {\it able not to} raise my hand') co-exist. A $QS$ is {\it compressed activity}, something which {\it is} and {\it is not} the case, {\it hic et nunc}. It cannot be thought in terms of identity but is expressed as a difference, as a {\it quantum of action}.

Our interpretation can be condensed in the following eight postulates. 

\begin{enumerate}

{\bf \item[I.] Hilbert Space:} QM is represented in a vector Hilbert space.

{\bf \item[II.] Potential State of Affairs (PSA):} A specific vector $\Psi$ with no given mathematical representation (basis) in Hilbert space represents a PSA; i.e., the definite existence of a multiplicity of {\it powers}, each one of them with a specific {\it potentia}.

{\bf \item[III.] Actual State of Affairs (ASA):} Given a PSA and the choice of a definite basis $B, B', B'',...,$ etc. ---or equivalently a Complete Set of Commuting Observables (C.S.C.O.)---, a context is defined in which a set of powers, each one of them with a definite potentia, are univocally determined as related to a specific experimental arrangement (which in turn corresponds to a definite ASA). The context builds a bridge between the potential and the actual realms, between quantum powers and classical objects. The experimental arrangement (in the ASA) allows the powers (in the PSA) to express themselves in actuality through elementary processes which produce {\it actual effectuations}.

{\bf \item[IV.] Quantum Situations, Powers and Potentia:} Given a PSA, $\Psi$, and the context or basis, we call a quantum situation to any superposition of one or more than one power. In general given the basis $B= \{ | \alpha_i \rangle \}$ the quantum situation $QS_{\Psi, B}$ is represented by the following superposition of powers:
\begin{equation}
c_{1} | \alpha_{1} \rangle + c_{2} | \alpha_{2} \rangle + ... + c_{n} | \alpha_{n} \rangle
\end{equation}

\noindent We write the quantum situation of the PSA, $\Psi$, in the context $B$ in terms of the order pair given by the elements of the basis and the coordinates in square modulus of the PSA in that basis:
\begin{equation}
QS_{\Psi, B} = (| \alpha_{i} \rangle, |c_{i}|^2)
\end{equation}

\noindent The elements of the basis, $| \alpha_{i} \rangle$, are interpreted in terms of {\it powers}. The coordinates of the elements of the basis, $|c_{i}|^2$, are interpreted as the {\it potentia} of the power $| \alpha_{i} \rangle$, respectively. Given the PSA and the context, the quantum situation, $QS_{\Psi, B}$, is univocally determined in terms of a set of powers and their respective potentia. (Notice that in contradistinction with the notion of {\it quantum state} the definition of a {\it quantum situation} is basis dependent.)

{\bf \item[V.] Elementary Process:} In QM we only observe discrete shifts of energy (quantum postulate). These discrete shifts are interpreted in terms of {\it elementary processes} which produce actual effectuations. An elementary process is the path which undertakes a power from the potential realm to its actual effectuation. This path is governed by the {\it immanent cause}\footnote{The {\it immanent cause} allows us to connect the power with its actual effectuation without destroying nor deteriorating the power itself. The immanent cause allows for the expression of effects remaining both in the effects and its cause. It does not only remain in itself in order to produce, but also, that which it produces stays within. Thus, in its production of effects the potential does not deteriorate by becoming actual ---as in the case of the hylomorphic scheme. Actual results are single effectuations, singularities which expose the superposition in the actual mode of existence, while superpositions remain evolving deterministically according to the Schr\"{o}dinger equation in the potential mode of existence, even interacting with other superpositions and producing new potential effectuations. } which allows the power to remain preexistent in the potential realm independently of its actual effectuation. Each power $| \alpha_{i} \rangle$ is univocally related to an elementary process represented by the projection operator $P_{\alpha_{i}} = | \alpha_{i} \rangle \langle \alpha_{i} |$.

{\bf \item[VI.] Actual Effectuation of Powers (Measurement):} Powers exist in the mode of being of ontological potentiality. An {\it actual effectuation} is the expression of a specific power in actuality. Different actual effectuations expose the different powers of a given $QS$. In order to learn about a specific PSA (constituted by a set of powers and their potentia) we must measure repeatedly the actual effectuations of each power exposed in the laboratory. (Notice that we consider a laboratory as constituted by the set of all possible experimental arrangements that can be related to the same $\Psi$.)

{\bf \item[VII.] Potentia (Born Rule):} A {\it potentia} is the strength of a power to exist in the potential realm and to express itself in the actual realm. Given a PSA, the potentia is represented via the Born rule. The potentia $p_{i}$ of the power $| \alpha_{i} \rangle$ in the specific PSA, $\Psi$, is given by:
\begin{equation}
Potentia \ (| \alpha_{i} \rangle, \Psi) = \langle \Psi | P_{\alpha_{i}} | \Psi \rangle = Tr[P_{ \Psi} P_{\alpha_{i}}]
\end{equation}

\noindent In order to learn about a $QS$ we must observe not only its powers (which are exposed in actuality through actual effectuations) but we must also measure the potentia of each respective power. In order to measure the potentia of each power we need to expose the $QS$ statistically through a repeated series of observations. The potentia, given by the Born rule, coincides with the probability frequency of repeated measurements when the number of observations goes to infinity.

{\bf \item[VIII.]  Potential Effectuation of Powers (Schr\"odinger Evolution):} Given a PSA, $\Psi$, powers and potentia evolve deterministically, independently of actual effectuations, producing {\it potential effectuations} according to the following unitary transformation:
\begin{equation}
i \hbar \frac{d}{dt} | \Psi (t) \rangle = H | \Psi (t) \rangle
\end{equation}
\noindent While {\it potential effectuations} evolve according to the Schr\"odinger equation, {\it actual effectuations} are particular expressions of each power (that constitutes the PSA, $\Psi$) in the actual realm. The ratio of such expressions in actuality is determined by the potentia of each power.
\end{enumerate}

\noindent According to our interpretation, just like classical physics talks about entities composed by properties that preexist in the actual realm, QM talks about powers with definite potentia that preexist in an ontological potential realm, independently of the specific actual context of inquiry or particular set of actualizations. This interpretational move allows us to define powers independently of the context regaining an objective picture of physical reality independent of measurements and subjective choices. The price we are willing to pay is the abandonment of the Newtonian metaphysical equation presupposed in the analysis of QM, exposing the fact that: Quantum Reality $\neq$ Actuality.

\section{Quantum Probability as an Objective Measure of the Potentia of Powers}

We would like to remark the fact that our notion of physical power is maybe the first physical notion to be characterized ontologically in terms of a probability measure. This concept escapes the ruling of actuality since it is  founded on a different set of metaphysical principles to that of classical entities. Indeed, powers are indetermined, paraconsistent and contextual existents. Powers can be superposed and entangled with different ---even contradictory--- powers \cite{daCostadeRonde13, daCostadeRonde15}. A power, contrary to a property which can be only {\it true} or {\it false} possesses an intrinsic probabilistic measure, namely, its potentia. A potentia is intrinsically statistical, but this statistical aspect has nothing to do with ignorance. It is instead an objective feature of quantum physical reality itself. 

Objective knowledge of properties is determined through yes-no experiments, one experiment is enough to completely characterize a property. Boolean classical logic and truth tables are suitable structures that allow us to define the elements of physical reality present in classical physics ---namely, actual properties. Contrary to classical properties, objective knowledge of powers with definite potentia can be only approached by performing statistical experiments. One experiment is simply not enough in order to determine the potentia of a power. Boolean classical logic and correspondence truth tables are not suitable structures and notions that would allow us to characterize the physical elements of reality present in QM which are, as we have discussed above, intrinsically statistical. Indeed, we must remark that our scheme implies the need of developing a new potential notion of truth, one that is not understood as a one-to-one relation of propositions with an actual state of affairs. We leave this work for future papers. Finally, we must remark that the ontological definition of the potentia of each power is determined already by the particular PSA, just in the same way a set of properties is determines in an ASA.

\section*{Conclusion}

Our conceptual physical scheme allows us to provide a metaphysical ground to Pauli's intuition ---also shared by Heisenberg, Popper, Margenau, Piron and many others--- that quantum statistics and quantum probability expose objective features of a quantum situation, instead of ``ignorance'' or ``inaccurate'' knowledge of an ASA. Quantum probabilities of physical quantities calculated though the Born rule become in our scheme the objective gnoseological counterpart of an ontological potentia, an element of physical reality that provides an objective measure of ontologically existent quantum powers. We believe that the key to disentangle the quantum riddle and create a coherent representation of quantum physical reality is to abandon the metaphysical dogma, presupposed in classical physics, that Reality = Actuality. We should acknowledge in this respect that ``common sense'' is just a name for the naturalization of dogmatic metaphysics. Indeed, our time calls to work on the elaboration of a new idea of reality, but in order not to engage ourselves in pseudoproblems we should also acknowledge right from the start that this project is in itself a metaphysical enterprise.

\section*{Acknowledgements} 

This work was partially supported by the following grants: FWO project G.0405.08 and FWO-research community W0.030.06. CONICET RES. 3646-14. I would like to thank G. Domenech and M. Graffigna for a careful reading of previous versions of this manuscript. Their many insightful comments and suggestions have been valuable for revising and improving our manuscript.

\end{document}